\documentclass[journal=jacsat,manuscript=article]{achemso}

\usepackage{chemformula} 
\usepackage[T1]{fontenc} 
\usepackage{siunitx}
\DeclareSIUnit\angstrom{\text {Å}}
\DeclareSIUnit\bar{bar}

\author{Michael Verhage}
\affiliation{Molecular Materials and Nanosystems (M2N) - Department of Applied Physics - Eindhoven University of Technology, Eindhoven, Netherlands}
\altaffiliation{Contributed equally}
\author{Emma van der Minne}
\affiliation{MESA+ Institute for Nanotechnology, Faculty of
Science and Technology, University of Twente, Enschede, Netherlands}
\altaffiliation{Contributed equally}
\author{Ellen M. Kiens}
\affiliation{MESA+ Institute for Nanotechnology, Faculty of
Science and Technology, University of Twente, Enschede, Netherlands} 
\author{Lucas Korol}
\affiliation{Department of Physics \& Engineering Physics, University of Saskatchewan, Saskatoon, Canada}
\author{Raymond J. Spiteri}
\affiliation{Department of Computer Science, University of Saskatchewan, Saskatoon, Canada}
\author{Gertjan Koster}
\affiliation{MESA+ Institute for Nanotechnology, Faculty of
Science and Technology, University of Twente, Enschede, Netherlands}
\author{Robert J. Green}
\affiliation{Stewart Blusson Quantum Matter Institute, University of British Columbia, Vancouver, Canada}
\alsoaffiliation{Department of Physics \& Engineering Physics, University of Saskatchewan, Saskatoon, Canada}
\author{Christoph Baeumer}
\affiliation{MESA+ Institute for Nanotechnology, Faculty of
Science and Technology, University of Twente, Enschede, Netherlands}
\alsoaffiliation{Peter Gruenberg Institute and JARA-FIT, Forschungszentrum Juelich GmbH, Juelich, Germany}
\email{c.baeumer@utwente.nl}
\author{Kees Flipse}
\affiliation{Molecular Materials and Nanosystems (M2N) - Department of Applied Physics - Eindhoven University of Technology, Eindhoven, Netherlands} 
\email{c.f.j.flipse@tue.nl}

\title
  {A complementary experimental study of epitaxial La$_{0.67}$Sr$_{0.33}$MnO$_3$ to identify morphological and chemical disorder}

\begin{document}

\begin{abstract}

Gaining insight into the characteristics of epitaxial complex oxide films is essential to control the behavior of devices and catalytic processes. It is known that substrate-induced strain, doping, and layer growth can affect the electronic and magnetic properties of the film's bulk. In this study, we demonstrate a clear distinction between the bulk and surface of thin films of La$_{0.67}$Sr$_{0.33}$MnO$_3$ in terms of chemical composition, electronic disorder, and surface morphology. We employed a combined experimental approach of X-ray based characterization methods and scanning probe microscopy. X-ray diffraction and resonant X-ray reflectivity revealed surface non-stochiometry in the strontium and lanthanum, as well as an accumulation of oxygen vacancies. Scanning tunneling microscopy showed a staggered growth surface morphology accompanied by an electronic phase separation (EPS) related to this non-stochiometry. The EPS is likely responsible for the temperature-dependent resistivity transition and is a cause of a proposed mixed-phase ferromagnetic and paramagnetic state near room temperature in these thin films.

\end{abstract}


\section{Introduction}

Manganite complex oxides are considered to be attractive materials for spin-based devices \cite{Yin2020SpinValves, Ogimoto2003TunnelingJunctions, Bruno2015InsightMapping} due to a half-metallic character \cite{Park1998DirectFerromagnet} and high spin polarization \cite{Bowen2003NearlyExperiments}. Doping the parent material LaMnO$_3$ with Sr in La$_{0.67}$Sr$_{0.33}$MnO$_3$ (LSMO) introduces insulator-to-metal and para-to-ferromagnetic transitions driven by complex interplay of structure, charge, and spin degrees of freedom \cite{Dagotto2001ColossalSeparation, Chmaissem2003Structural3}.  LSMO possesses an advantageous Curie temperature of \SI{370}{\kelvin} \cite{Urushibara1995Insulatormetal3} that potentially enables spin control at room temperature. The control of these properties is of interest in fundamental electrochemical studies \cite{VanDerMinne2023DomainSplitting, Xu2023MagneticCatalyst} of manganites. LSMO has been used as a model catalyst to study the oxygen reduction and evolution reactions (ORR and OER, respectively) due to the tunability of the active manganese site and its promising bifunctional activity \cite{Scholz2016RotatingConcentration, Lee2021ContributionHeterostructures, Gonell2020StructureElectrocatalysis, Hale2022ThermodynamicStudy}. A recent experiment has shown that active strain can be used to regulate the activity of LSMO thin films in the OER \cite{Qi2022StrainFilms}. This, combined with its high spin polarization, could be a useful tool for exploring the influence of electron spin on the OER \cite{Gracia2017SpinElectrochemistry}.

In addition to catalysis, and closely related to spin, LSMO has been extensively studied for phenomena such as colossal magnetoresistance \cite{Moreo2000GiantCompounds, Burgy2001ColossalInhomogeneities} and electroresistance \cite{Debnath2003CurrentinducedFilms}. These phenomena are understood within the framework of electronic phase separation (EPS) and magnetic phase separation \cite{Miao2020DirectManganites}. Reports of chemical \cite{Zhu2016ChemicalManganites, Miao2020DirectManganites} and pressure \cite{Baldini2015OriginManganite} induced EPS of manganites are associated with bulk disorder and are essential to this family of metal oxides. A great deal of research has been done on LSMO thin films, exploring how changes in parameters such as substrate-induced strain (tensile or compressive), film thickness, substrate termination, \cite{Lopez-Mir2016GrowthFilms} and growth conditions affect structural, electronic, and magnetic behavior \cite{Zhang2021LargeFilms, Yin2018LargeFilms, Liu2016SubstraterelatedFilms, Carreira2019NanoscaleInterfaces}. However, there are still insights to be gained in the discrepancy between the surface structure and chemical composition and that of the bulk. To address this, we conducted both nanoscale imaging and spectroscopic analysis on a LSMO thin film to link nanoscale surface characteristics \cite{Carreira2019NanoscaleInterfaces} with those of the bulk to gain a comprehensive view of our LSMO films. 

This work presents evidence that, for LSMO epitaxial thin films, both structural and chemical disorder persist at the surface, as revealed by high-resolution scanning probe microscopy (SPM) imaging. We observe nanometer-scale corrugation of the surface \cite{Gambardella2014SurfaceFilms}, consisting of the formation of a staggered growth of small islands \cite{Tselev2015SurfacePressure}, that deviates from an atomically smooth interpretation. We define an atomically smooth film as a surface roughness variation on the scale of atomic sizes with a correlation length on the order of the widths of the vicinal steps. Pandya \textit{et al.} \cite{Pandya2016StraininducedFilms} have reported the emergence of such staggered growth or ``wedding cake" structures, which are controlled by substrate-induced strain. Similarly, Kelley \textit{et al.} \cite{Kelley2021ExoticFilms} have observed these islands in LSMO. 

Furthermore, we used area-averaged chemical mapping by resonant X-ray reflectometry (RXR) to investigate a gradient in the chemical composition from a stochiometric bulk to an excess of the Sr dopant near the surface. The chemical disorder is accompanied by a partial change in the valence state of Mn from Mn$^{3+}$ to Mn$^{2+}$ and an increase in oxygen vacancies \cite{DeJong2005EvidenceFilms}. Sr segregation in the surface region was previously identified in Ref. \cite{Dulli2000Surface3} and linked to morphology restructuring \cite{Gambardella2014SurfaceFilms}. 

Using scanning tunneling spectroscopy (STS), we detect an EPS across the surface that is related to the nature of the staggered growth and chemical disorder. This relation between rough surface morphology and EPS has been reported for Ca-doped LaCaMnO$_3$ (LCMO) films \cite{Kelly2009CorrelationsSTM}, and thus appears to be a common feature of the manganite perovskite family.

The EPS is likely to be the cause of a magnetic phase separation in LSMO. Previous studies of Ca-doped manganites have reported the presence of ferromagnetic (FM) and paramagnetic (PM) areas that coexist and lead to percolation \cite{Baldini2015OriginManganite, Li2002CompetitionManganites}. By studying the temperature-dependent resistance, a phenomenological model of phase separation was constructed between the FM and PM states \cite{Li2002CompetitionManganites}, with a dominant FM phase at lower temperatures and a mixture of FM and PM at higher temperatures. A deeper understanding of the LSMO surface enables insights into its applications in devices and catalysis. 

\section{Results and Discussion}

\begin{figure*}
\centering
        \includegraphics[width=\columnwidth]{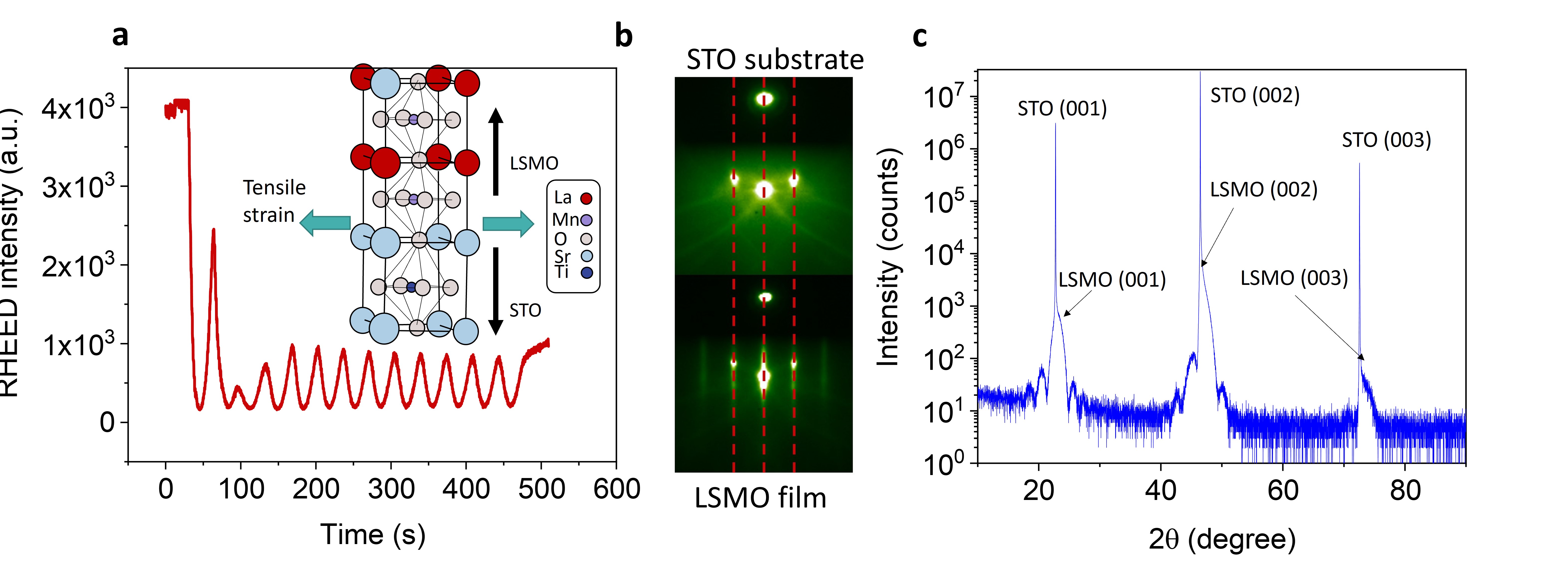}
            \caption{\textbf{In-situ growth characterization and structure of 13 u.c. LSMO films.} (\textbf{a}) In-situ RHEED intensity oscillations for a 13 unit cell (u.c.) thick LSMO film grown on STO. Inset: schematic illustration of the atomic structure. (\textbf{b}) RHEED patterns of the substrate (top) and the film (bottom). The red dashed lines are a guide for the eye and indicate that the distance between the diffraction spots does not change. (\textbf{c}) Wide angle $2\Theta$-$\omega$ scan, indicating single (001)-oriented LSMO. From the position of the LSMO 002 peak a c-axis parameter of \SI{0.385}{\nano\meter} is obtained. }
            \label{fig:Growth_Characterisation}
\end{figure*}

LSMO films of \SI{13}{} unit cell (u.c.) thickness were grown with pulsed laser deposition (PLD) on SrTiO$_3$ (STO) 001 substrates; further experimental details are described in \textit{Methods}. We use reflection high-energy electron diffraction (RHEED) to monitor growth \textit{in situ}. The presence of clear RHEED intensity oscillations indicates layer-by-layer growth, as can be observed in \textbf{Fig. \ref{fig:Growth_Characterisation}a}. A schematic illustration of the crystal structure is shown in the inset of \textbf{Fig. \ref{fig:Growth_Characterisation}a}, and the growth on STO introduces tensile strain. A decrease in peak amplitude over time and faint streaks observed along with sharp diffraction spots in \textbf{Fig. \ref{fig:Growth_Characterisation}b} point toward the presence of disorder and roughness \cite{Koster1999ArtificiallyDeposition} of the LSMO surface. The equal distance between the diffraction spots in the RHEED pattern, \textbf{Fig. \ref{fig:Growth_Characterisation}b}, indicates that the in-plane lattice parameters of the film match those of the substrate, implying the presence of tensile strain \cite{Yang2010StrainFilms}. The film quality was further verified by X-ray diffraction (XRD), indicating a pure (001) oriented LSMO phase as seen in \textbf{Fig. \ref{fig:Growth_Characterisation}c}.


\begin{figure*}
    \centering
        \includegraphics[width=\columnwidth]{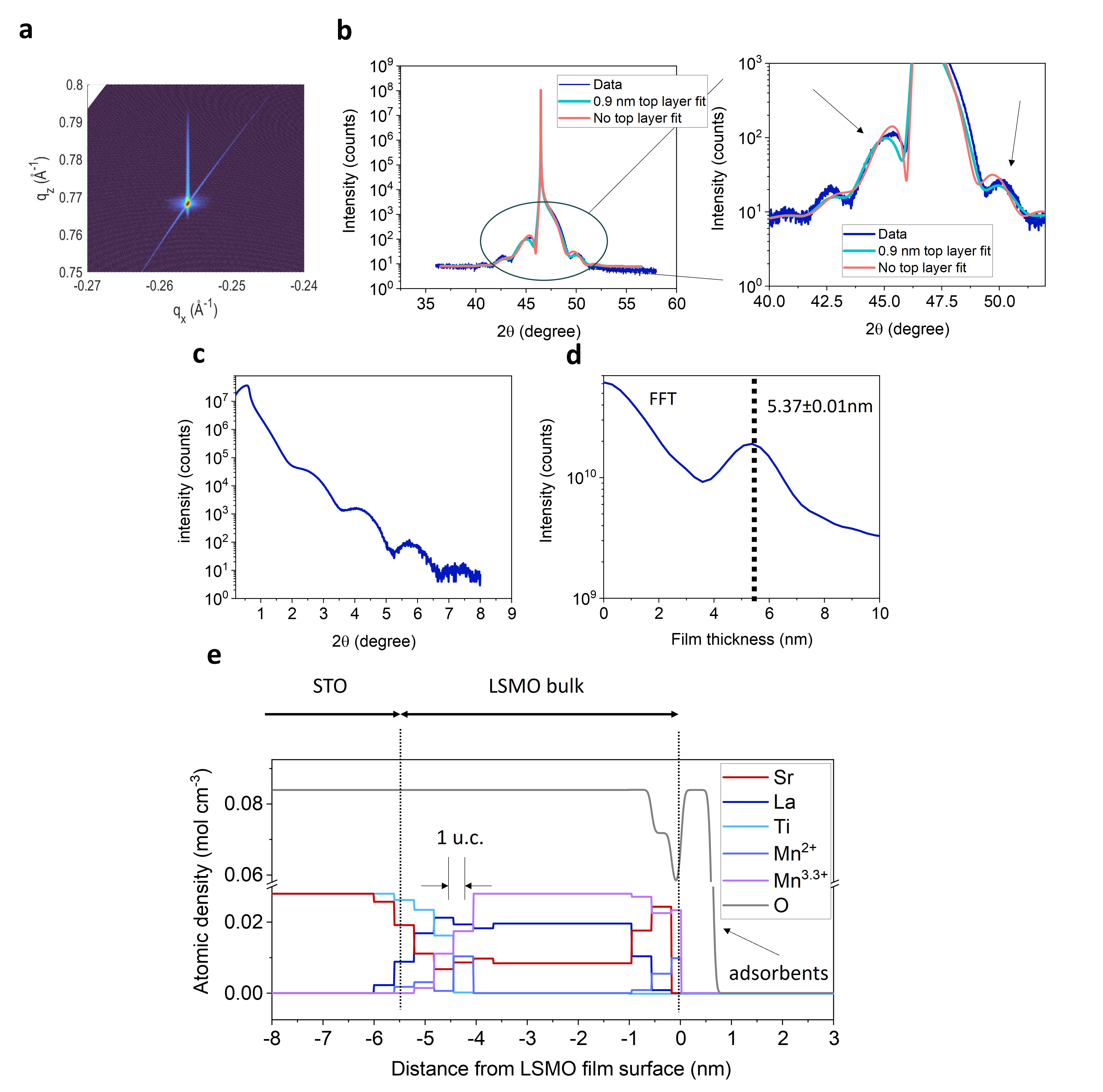}
            \caption{\textbf{XRD and RXR analysis of the \SI{13} u.c. LSMO film.} (\textbf{a}) Reciprocal space map around the STO 103 peak. (\textbf{b}) XRD $2\Theta$-$\omega$ scans around the STO 002 peak along with dynamical simulated diffactrograms. The simulations were performed without or with a lower crystalline layer of \SI{0.9}{\nano\meter}. The large deviations of the fringe at 42$^\circ$ are probably induced by the broadening of the experimental substrate peak, which is not taken into account in the simulations. Simulations were done with the Stephanov X-ray server \cite{Stepanov2004XrayScattering}. The arrows indicate large deviations from the data for the simulation without a top layer. (\textbf{c}) XRR data for 13 u.c. LSMO film. (\textbf{d}) The Fourier transform of \textbf{c}, which indicates an LSMO film thickness of 5.37±0.01 nm. (\textbf{e}) Atomic depth profiles of La, Ti, Sr, Mn$^{2+}$, Mn$^{3.3+}$ and O obtained using resonant X-ray reflectivity (RXR) at 300K.}
            \label{fig:xrd_Charactersiation}
\end{figure*}

\begin{figure*}
    \centering
        \includegraphics[scale=0.6]{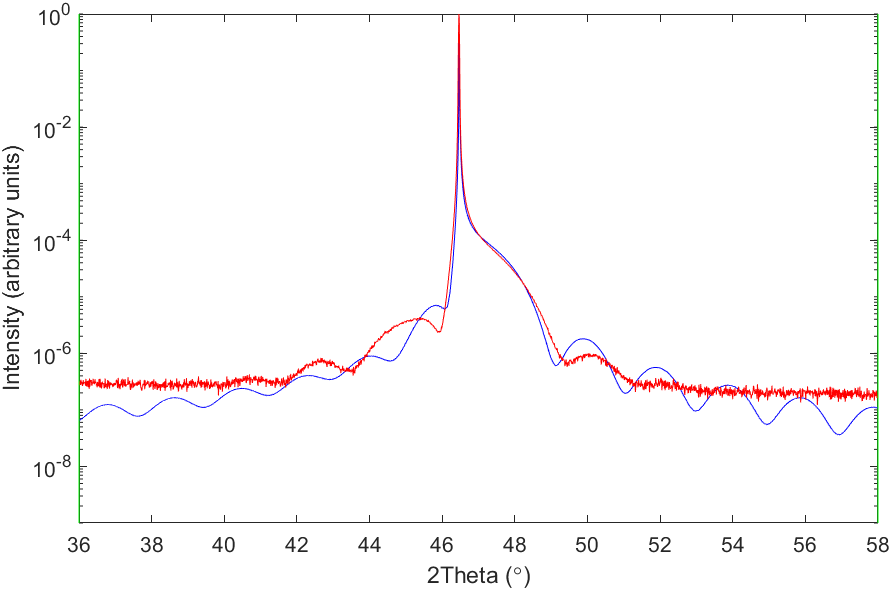}
            \caption{\textbf{Comparison of the measured and expected diffractogram  of the LSMO film peaks around the STO 002 peak in a $2\Theta$-$\omega$ scan}. The expected diffractrogram is obtained with the Interactive XRDFit software \cite{Lichtensteiger2018InteractiveXRDFitHeterostructures} using a thickness of 13 u.c. and c = 3.845 $\si{\angstrom}$.}
            \label{fig:quickfit}
\end{figure*}

Further XRD analysis on the LSMO 002 peaks was performed to investigate the crystallinity and strain state of the thin film. The reciprocal space map around the STO 103 peak in \textbf{Fig. \ref{fig:xrd_Charactersiation}a} confirms that the bulk of the film exhibits a coherent tensile strain. The clear Laue fringes in \textbf{Fig. \ref{fig:xrd_Charactersiation}b} indicate good crystallinity in the bulk of the film. However, comparing these measured fringes with an expected diffractogram of a 13 u.c. thick film (\textbf{S1 - Fig. \ref{fig:quickfit}}) indicates a measured film thickness below 13 u.c. Furthermore, the measured crystalline layer is thinner than the thickness obtained from the Fourier transform of the XRR data in \textbf{Fig. \ref{fig:xrd_Charactersiation}c, d} of 5.37±0.01 nm, which is close to the expected value for a 13 u.c. thin film. This indicates, as discussed in more detail below, that a layer of low crystallinity, but with an electronic density similar to that of the LSMO bulk, is present. See \textit{Methods, XRD} for the modeling details of the crystallinity variability in the film. Correlating this with the resonant X-ray reflectivity (RXR) results, see below, indicates that this layer of lower crystallinity is positioned at or near the surface. The existence of such a layer is further confirmed by the asymmetric period around the STO 002 peak, which indicates asymmetry in the film. The data can be better fitted when a layer of approximately \SI{0.9}{\nano\meter} with low crystallinity is considered on the film surface, as shown in \textbf{Fig. \ref{fig:xrd_Charactersiation}b}. 

Due to the crystalline dissimilarity between the bulk and the surface, we turned to RXR to study the chemical disorder \cite{Green2020ResonantHeterostructures}. With RXR, subnanometer depth-resolution of the elemental composition can be obtained. \cite{Hamann-Borrero2016ValencestateHeterointerfaces}. The depth profile of the LSMO film shown in \textbf{Fig. \ref{fig:xrd_Charactersiation}e} indicates non-stochiometric regions near the substrate/film interface and the film surface. At the buried interface, the intermixing of the STO substrate with the LSMO film persists for more than \SI{1}{\nano\meter}, inducing a La:Sr ratio that deviates from the stochiometric value of 2:1. The non-stochiometry at the buried interface is a likely cause of the reported magnetic dead layers of LSMO \cite{Porter2017MagneticRevisited}. Furthermore, we observe Mn$^{2+}$ at the buried interface, which could be an effect of the intermixing of Ti$^{4+}$ atoms. Near the surface, we observe Sr segregation and La deficiency and the formation of Mn$^{2+}$ species. The latter could be induced by the presence of oxygen vacancies in the surface layer \cite{DeJong2005EvidenceFilms, Li2012InterfaceFilms}. Surface non-stochiometry is unlikely to be introduced by beam damage because with repeating scans no changes in stochiometry were noted. On the basis of these results, the non-stochiometry in the surface of the film could be a reason for the lower crystallinity in the surface compared to that of the bulk.

\subsection{Surface imaging}

\begin{figure*}
    \centering
        \includegraphics[scale=0.5]{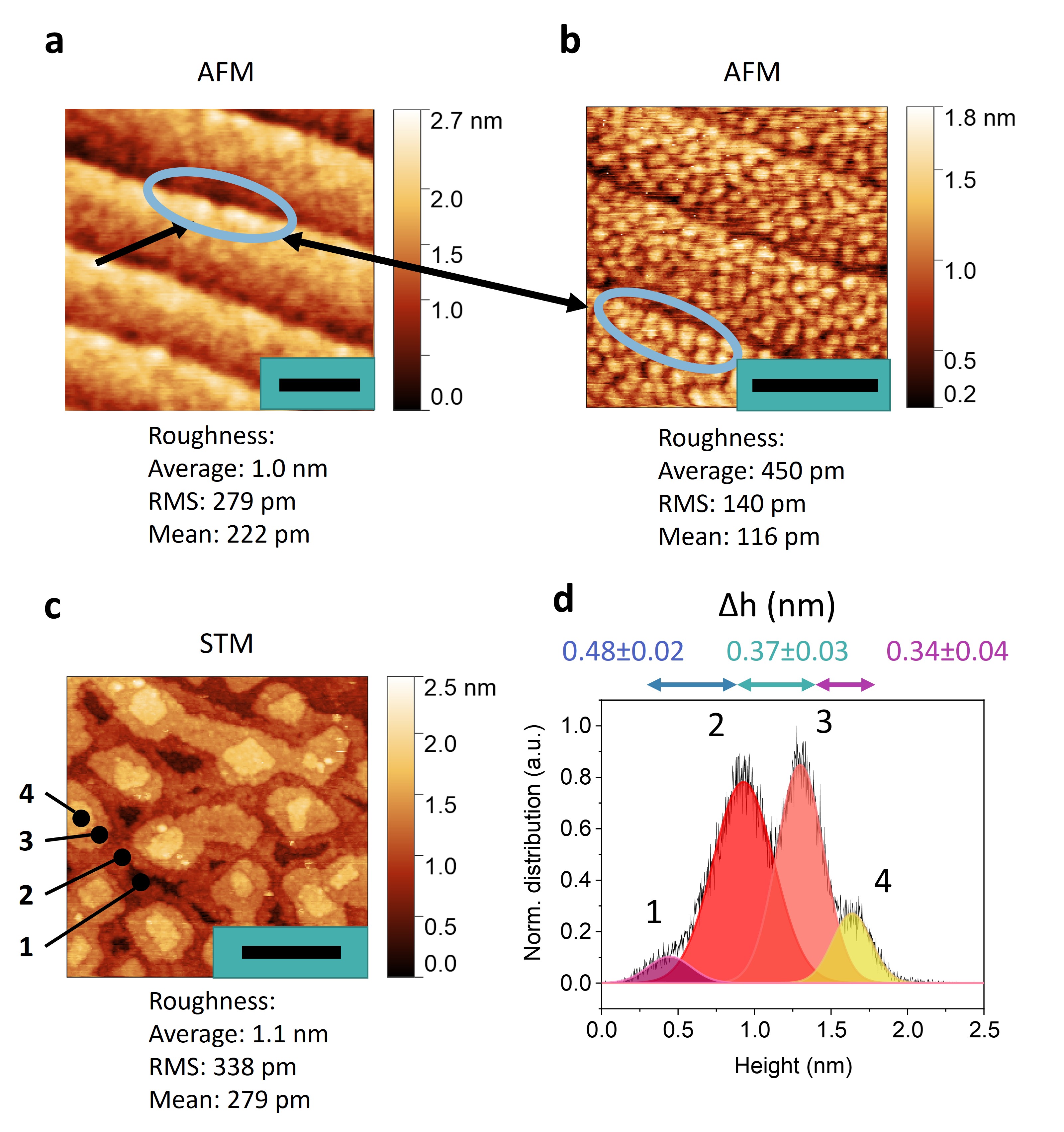}
            \caption{\textbf{Surface structure of 13 u.c. LSMO thin films.} (\textbf{a}) AFM imaging with a \SI{30}{\nano\meter} radius tip showing mostly contrast from the step edges. The scale bar is equal to \SI{100}{\nano\meter}. (\textbf{b}) AFM image with a tip radius of \SI{5}{\nano\meter} indicating a granular surface morphology. Scale bar equals \SI{40}{\nano\meter}. (\textbf{c}) High resolution STM image (\SI{800}{\milli\volt}, \SI{100}{\pico\ampere}) revealing a staggered growth. Numbers indicate consecutive layers. The scale bar is equal to \SI{10}{\nano\meter}. (\textbf{d})  Height distribution of \textbf{c} with the peak spacing indicating the stacking of layers.}
            \label{fig:Surface_Structure}
\end{figure*}

\begin{figure*}
    \centering
        \includegraphics[scale=0.6]{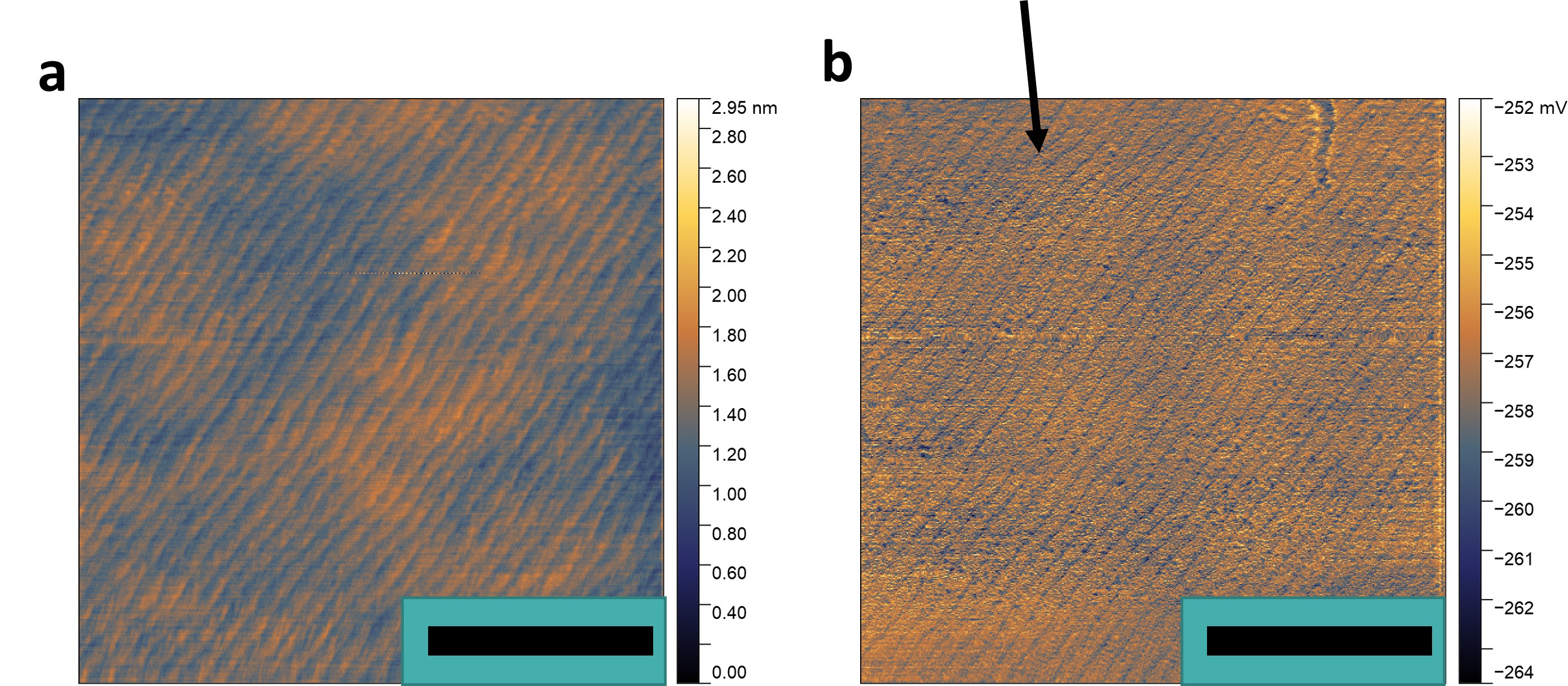}
            \caption{\textbf{Contact-mode AFM.} (\textbf{a}) Topographic map with contact-mode AFM showing contrast predominantly from the step edges. (\textbf{b}) Friction map shows stronger signal (blue) indicated along the step edges (indicated by the black arrow), while it shows little on the plateaus.}
            \label{fig:friction}
\end{figure*}

\begin{figure*}
    \centering
        \includegraphics[scale=0.6]{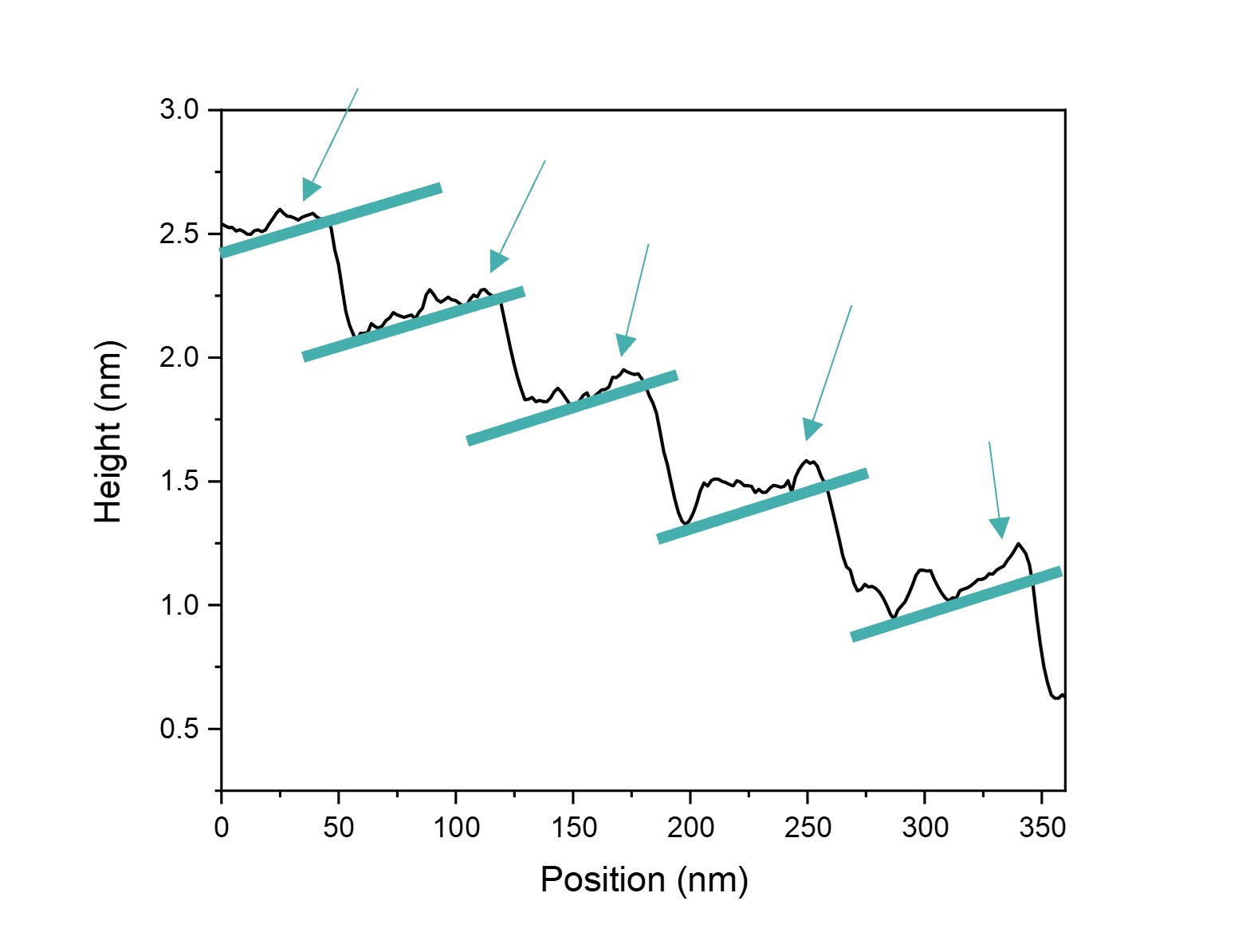}
            \caption{\textbf{Line cross-section of Fig. \ref{fig:Surface_Structure}a on the main text.} Green arrows indicate upward bulging of the topography near the step edge. The plateau's are indicated with the green lines.}
            \label{fig:Cross_Section}
\end{figure*}

From the X-ray analysis, we identified that the surface of our LSMO films deviates from the high-crystalline, chemically stochiometric bulk. We turned to atomic force microscopy (AFM) to study the surface structure of 13 u.c. LSMO films on STO in further detail. First, we imaged the surface with a rather large diameter tip. Our custom AFM probe has a pyramidal tip \cite{Verhage2023SwitchablemagnetisationSensor} with a radius of around \SI{30}{\nano\meter}, similar to commercial Si cantilevers. We used Frequency-Modulation (FM-AFM) feedback in repulsive mode. In this mode, the tip gently touches the sample and is lifted again during each oscillation cycle, similar to tapping-mode AFM. The benefit of this approach is that the AFM is most sensitive to short-range forces, increasing the resolution \cite{Giessibl2019QPlusMicroscope}. However, the obtainable resolution is mainly limited by the radius of the tip \cite{Heath2021LocalizationMicroscopy}. See \textit{Methods} for further experimental details. As expected from the vicinal cut STO, the epitaxial thin film LSMO demonstrated flat plateaus and step edges, \textbf{ Fig. \ref{fig:Surface_Structure}a}.  These stepped film surfaces have often been reported in the literature for a wide range of transition metal oxide perovskites and layer thicknesses \cite{Boschker2011OptimizedCharacteristics,Molinari2022TailoringEngineering}. The steps give rise to significant contrast in the AFM due to the convolution of the tip and the lateral friction when the side of the tip hits the atomic edge \cite{Chen2019ChemicalSteps}. To show that this contrast emerges from repulsive force or friction, even in FM-AFM when the side of the tip hits the step edge, we performed contact AFM. The topographic results are given in \textbf{S2 Fig. \ref{fig:friction}}, where a clear contrast near the edge of the step is observed. Hence, step edges can dominate the AFM signal due to tip-sample friction during contact and morphological details of the plateau surfaces can be overlooked. The conformation of the surface is slightly bulged upward near the step edge (\textbf{S3 Fig. \ref{fig:Cross_Section}}). However, with the resolution obtained here, only a faint hint at nanometer-scale features of the scale of less than 1 u.c. height variation could be identified as highlighted with the black arrows.

In order to improve lateral imaging resolution, we switched to a diamond tip with a radius of \SI{5}{\nano\meter}; see \textit{Methods}. \textbf{Fig. \ref{fig:Surface_Structure}b} shows a completely different surface morphology, despite it being the same sample as in \textbf{Fig. \ref{fig:Surface_Structure}a}. The surface is densely populated with semispherical objects with an average roughness of \SI{450}{\pico\meter} and a mean roughness ($S_a$) of \SI{140}{\pico\meter}. These features are more closely packed and slightly taller near the step edge, as highlighted in blue, which explains why AFM with a lower imaging resolution of \textbf{Fig. \ref{fig:Surface_Structure}a} showed bulging towards the step edge.

To further explore the nanoscale surface structures of the LSMO films, we used scanning tunneling microscopy (STM) in a UHV-SPM setup. The tip had a radius of less than \SI{1}{\nano\meter} and the STM was mainly sensitive to electronic or height variations on the atomic scale. The STM image in \textbf{Fig. \ref{fig:Surface_Structure}c} showed that the semispherical structures are actually small stacked islands. These morphologies are known as staggered growth, or “wedding cake” islands \cite{Pandya2016StraininducedFilms}. Similar features were observed for Ca-doped manganite \cite{Kelly2009CorrelationsSTM} and are termed mounds \cite{Tselev2015SurfacePressure}. Our tip-radius-dependent imaging highlights that, without using ultra-sharp SPM tips, nanoscale features may be hidden and the surface morphology inaccurately interpreted as atomically smooth on a scale of the width of the STO plateaus.

In \textbf{Fig. \ref{fig:Surface_Structure}c}, the most deeply buried layer, as indicated by the number $1$, is coalesced with only a few voids left in the area. We observe up to \SI{3}{} layers stacked on top of this coalesced layer in the form of "wedding cake" islands or staggered growth, as shown numerically in the figure. By analyzing the statistical distribution of the height data, \textbf{Fig. \ref{fig:Surface_Structure}d}, we observe a combination of \SI{4}{} height distributions. Each peak corresponds to a higher positioned layer in \textbf{Fig. \ref{fig:Surface_Structure}c}. We conducted a Gaussian fit to identify the peaks, as indicated by the colored curves. These peaks are not evenly spaced, as indicated by the numbers corresponding to the increase in height ($\Delta h$) of each stacking layer. However, they are close to the values for the thickness of 1 u.c. (\SI{0.385}{\nano\meter}) except for the layer numbered $1$ to $2$. This could be due to the fact that the area of the layer $1$ is almost completely coalesced, so the convolution of the tip in the few small gaps has a negative effect on the accurate measurement of the thickness of the film. In this work, we observe clear layer-by-layer growth but no half-integer heights, which is different from the work of Ref. \cite{Tselev2015SurfacePressure}, where a peak spacing around \SI{200}{\pico\meter} was observed indicating a mix of surface termination growth.

The variation in the peak-to-peak height of staggered growth can be as large as \SI{2.5} {\nano\meter}, as can be seen in the maximum height value of \textbf{Fig. \ref{fig:Surface_Structure}c}. These large heights originate from individual dispersed spherical-like features across the surface. The film has an average roughness of \SI{1.1}{\nano\meter} and an RMS roughness of \SI{338}{\pico\meter}. With a film thickness of \SI{13}{} u.c., this average roughness accounts for approximately \SI{20}{\percent} of the total film thickness. Based on the XRD results of \textbf{Fig. \ref{fig:xrd_Charactersiation}}, we observed that the thickness of the surface layer with low crystallinity in the \SI{13}{} u.c. LSMO film was approximately \SI{0.9}{\nano\meter}. This corresponds rather well with the observed staggered growth average film roughness by the STM and the off-stochiomtery of 2.5 u.c. ($\sim$\SI{1}{\nano\meter}) from RXR. 

\begin{figure*}
    \centering
        \includegraphics[scale=0.5]{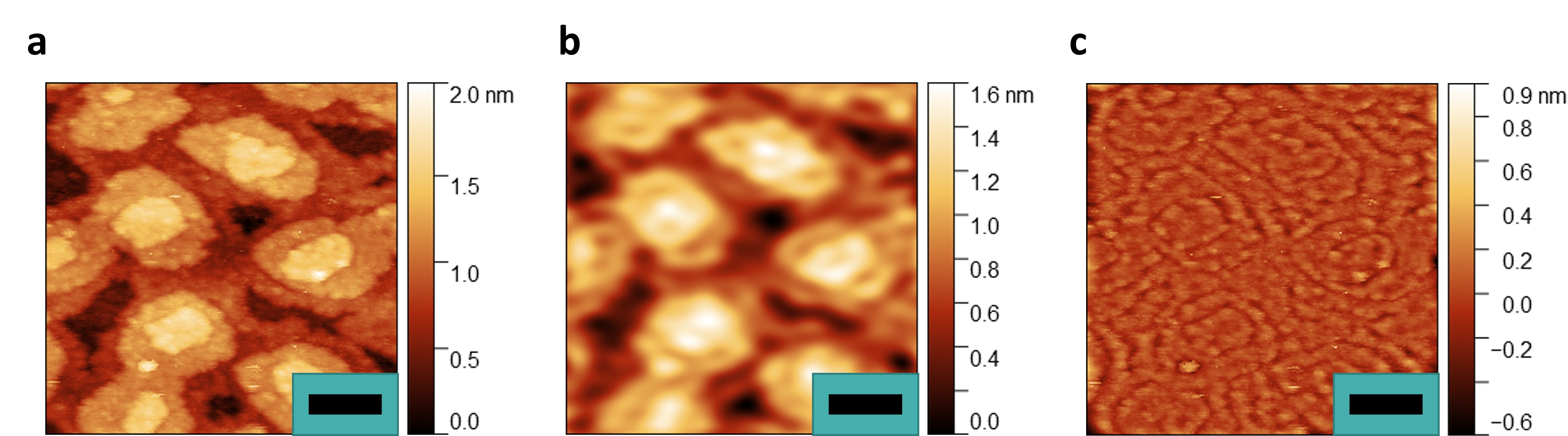}
            \caption{\textbf{FFT filtering showing spherical features across the staggered growth of the LSMO surface.}
            (\textbf{a}) STM image of \SI{13} u.c. LSMO thin film. (\textbf{b}) Fast Fourier filtering of the main staggered growth. (\textbf{c}) After the removal of the staggered growth by FFT filtering leaves only the spherical features of the film surface highlighted.}
            \label{fig:FFT_grains}
\end{figure*}

\begin{figure*}
    \centering
        \includegraphics[scale=0.6]{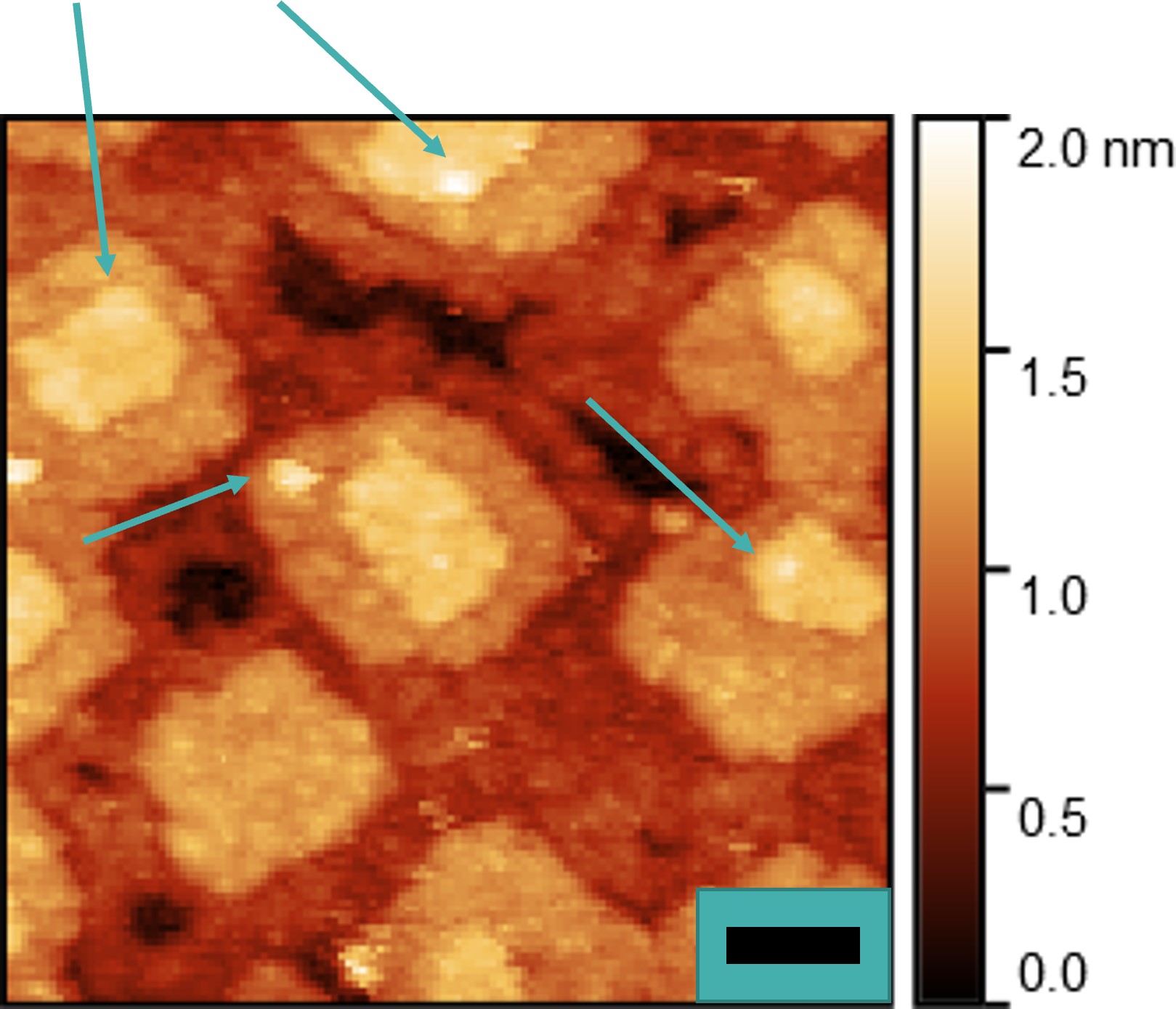}
            \caption{\textbf{STM observation of individual spherical features.} STM image of \SI{13}{} u.c. LSMO/STO film with staggered growth showing individual spherical features distributed ontop of the upper layers, as indicated with the arrows.}
            \label{fig:Top_spherical particles}
\end{figure*}

To relate the observed surface structures to bulk crystallinity, we propose a two-step growth process. First, the deposited species during PLD nucleate and grow, manifesting as spherical features as observed in the SPM images. These spherical features move across the surface by diffusion and lateral aggregation but are impeded near the edges of the vicinal step due to the Erlich-Swoebel (ES) barrier \cite{Pandya2016StraininducedFilms}. This barrier near the step edge impedes downward diffusion of surface features, leading to a densely packed morphology of islands toward the step edge region. Consequently, just below the step edge, there is a material deficiency, and few staggered structures can form there. We argue that the spherical features themselves are not necessarily an integer number of atomic u.c. stackings of the ABO$_3$ perovskite. However, after embedding the features into a staggered growth layer, the thickness converges to integer u.c. heights (\textbf{Fig. \ref{fig:Surface_Structure}d}), likely driven by a crystal restructuring. The disorder in the morphology of the staggered growth breaks long-range crystallinity and explains why a thinner fully crystalline bulk thickness is obtained in comparison to the total film thickness by XRD fitting. Possibly, Sr segregation or oxygen vacancies \cite{Li2012InterfaceFilms} break the stochiometery of the islands, as previously observed with RXR in \textbf{Fig. \ref{fig:xrd_Charactersiation}e}. The spherical nature of the islands is further identified by FFT filtering of the STM data in \textbf{S4 Fig. \ref{fig:FFT_grains}}. When features associated with the staggered growth are removed, the spherical features remain and can be clearly observed. During PLD, these spherical features are formed on top of the islands and experience another ES barrier at the periphery of the island below.  The coalescence and growth of approximately circular islands are naturally accompanied by gaps across layers. As the islands expand in width, the gaps gradually coalesce. Only when these islands have reached a sufficient size can second layer nucleation occur \cite{Beausoleil2014ImpactSurfaces}. Consequently, we did not observe nucleation of smaller islands or spherical features on nearly all of the highest islands. Rather rare in presence, we sometimes observed small individual spherical features scattered throughout the surface, as seen in \textbf{S5 Fig. \ref{fig:Top_spherical particles}}. This observation is consistent with the work of Tershoff \textit{et al.} \cite{Tersoff1994CriticalGrowth}. In their work, it was shown that nucleation of a second layer occurs only when an island has reached a critical diameter in combination with a sufficiently strong ES barrier of the island boundaries. This seems to imply that before a critical size is reached, surface species can diffuse through the ES barrier and contribute to an increase in the lateral dimension of the islands. Only when a critical size has been reached, the ES barrier is significant in magnitude to block species that diffuse downward and form a new layer of staggered growth. The second step of the film growth process occurs after the coalescing of a fully closed layer. We argue that reconstruction can occur in these closed layers, forming a highly ordered crystalline bulk, as seen in the XRD data. However, local defects can still persist \cite{Tselev2015SurfacePressure}, leading to a chemical dopant-induced disorder \cite{Miao2020DirectManganites}. 

\subsection{Variation in surface electronic structure}
\begin{figure*}[b!]
    \centering
        \includegraphics[scale=0.5]{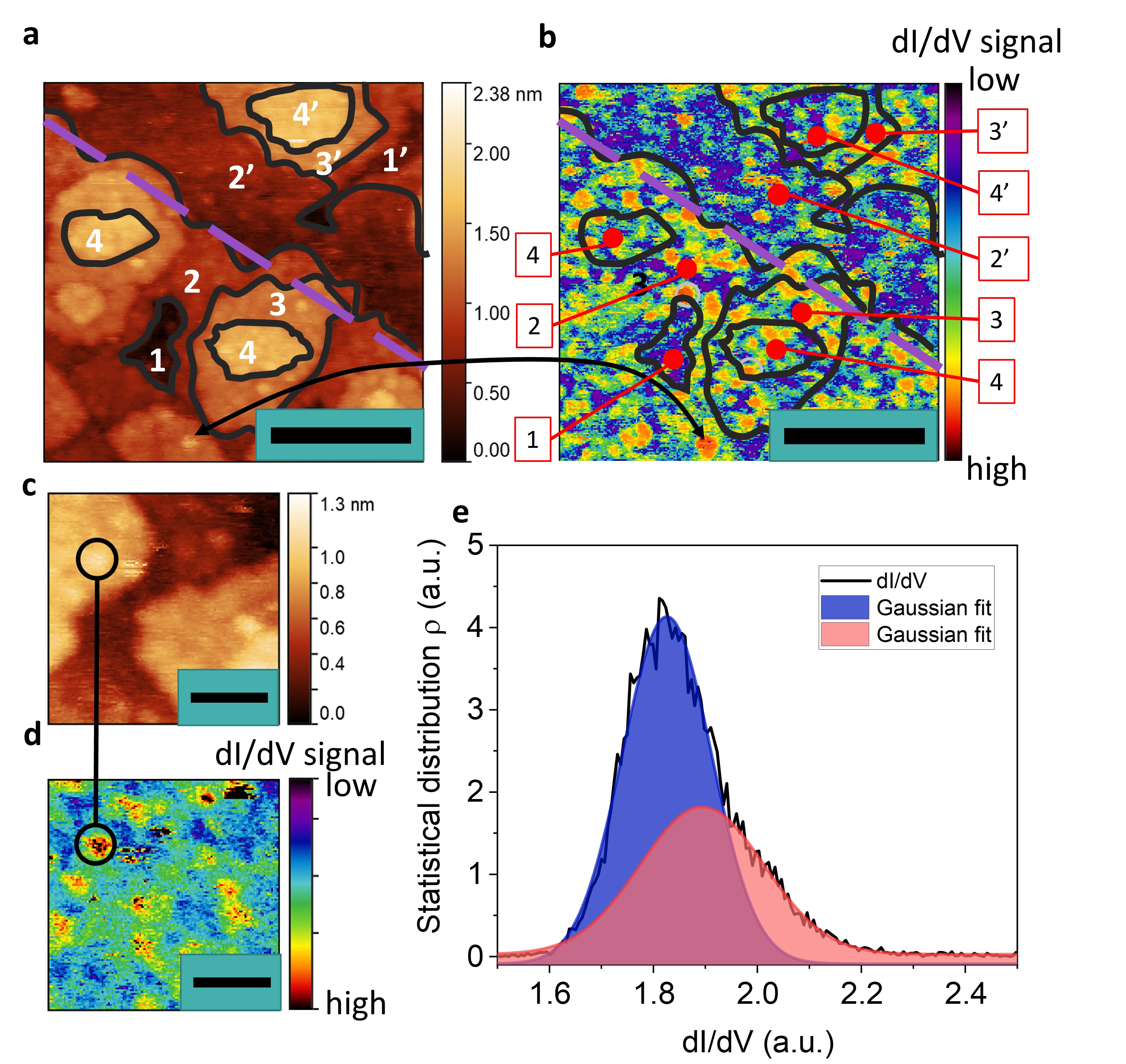}
            \caption{\textbf{Scanning tunneling spectroscopy on 13 u.c. LSMO thin films demonstrating electronic phase separation.} (\textbf{a}) Topography STM image (\SI{800}{\milli\volt}, \SI{100}{\pico\ampere}) with numbers indicating staggered growth islands. The purple dashed line indicates the step edge.  The numbers $1' - 4'$ are layers located on an STO plateau below the step edge. The scale bar is equal to \SI{10}{\nano\meter}. (\textbf{b}) Corresponding dI/dV map at \SI{800}{\milli\volt}. The scale bar is equal to \SI{10}{\nano\meter}. (\textbf{c}) Small-scale topography (\SI{800}{\milli\volt}, \SI{100}{\pico\ampere}) and (\textbf{d}) corresponding dI/dV map. The scale bar is equal to \SI{2}{\nano\meter}. The black circle highlights the increased conductivity of a circular feature observed in the topography. (\textbf{e}) Statistical dI/dV distribution of \textbf{d} fitted with a Gaussian and showing two distributions of LDOS.}
            \label{fig:STM_STS}
\end{figure*}

The local electronic structure of the complex morphology of the LSMO surface was studied in more depth. The presence of electronic disorder or EPS was investigated with scanning tunneling microscopy (STM) and tunneling spectroscopy (STS), respectively. \textbf{Figs. \ref{fig:STM_STS}a} and \textbf{\ref{fig:STM_STS}b} show the topography and simultaneously mapped dI/dV or local density of states (LDOS) measured at \SI{800}{\milli\volt} at \SI{300}{\kelvin}, respectively. For the dI/dV map, the areas in red have a larger LDOS than the areas in blue. Staggered growth island contours are highlighted with black lines and numbered at each consecutive stacked layer. From the dI/dV map, we observe that the layer, indicated with \textbf{\SI{1}{}} in \textbf{Fig. \ref{fig:STM_STS}b} is rather homogeneous in LDOS with only some smaller local LDOS spots. This layer is almost structurally coalesced with few remaining voids. However, the staggered growth layers \SI{2}{} - \SI{4}{} show a greater variation in the intensity of LDOS. The dI/dV signal can be correlated with the presence of surface spherical features, as highlighted by the black circles in \textbf{Figs. \ref{fig:STM_STS}c, d}. These spherical features possess a larger LDOS, and hence it is likely that they possess a non-stochiometric chemical composition, with an excess of Sr making the spherical features more conductive. These spherical features are rather dispersed throughout the staggered growth.   

The EPS persists down to the nanometer scale. In \textbf{Figs. \ref{fig:STM_STS}c} and \textbf{\ref{fig:STM_STS}d} a \SI{10}{}$\times$\SI{10}{\nano\meter} image reveals a clear inhomogeneity of dI/dV with some electronic domains less than \SI{1}{\nano\meter} in diameter. During growth, Sr adatoms may diffuse through the staggered growth before stochiometry is achieved in the bulk. The surface roughness accompanying EPS has previously been reported for Ca-doped LMO \cite{Kelly2009CorrelationsSTM}. Below the step edge, as indicated by the purple dotted line in width \textbf{Figs. \ref{fig:STM_STS}a, b}, a small region numbered $2'$, of less than \SI{10}{\nano\meter} in width is observed. It has little staggered growth and rather uniform LDOS, showing that surface morphology and the EPS are correlated for the LSMO thin films. 

\begin{figure*}[b!]
    \centering
        \includegraphics[width=\columnwidth]{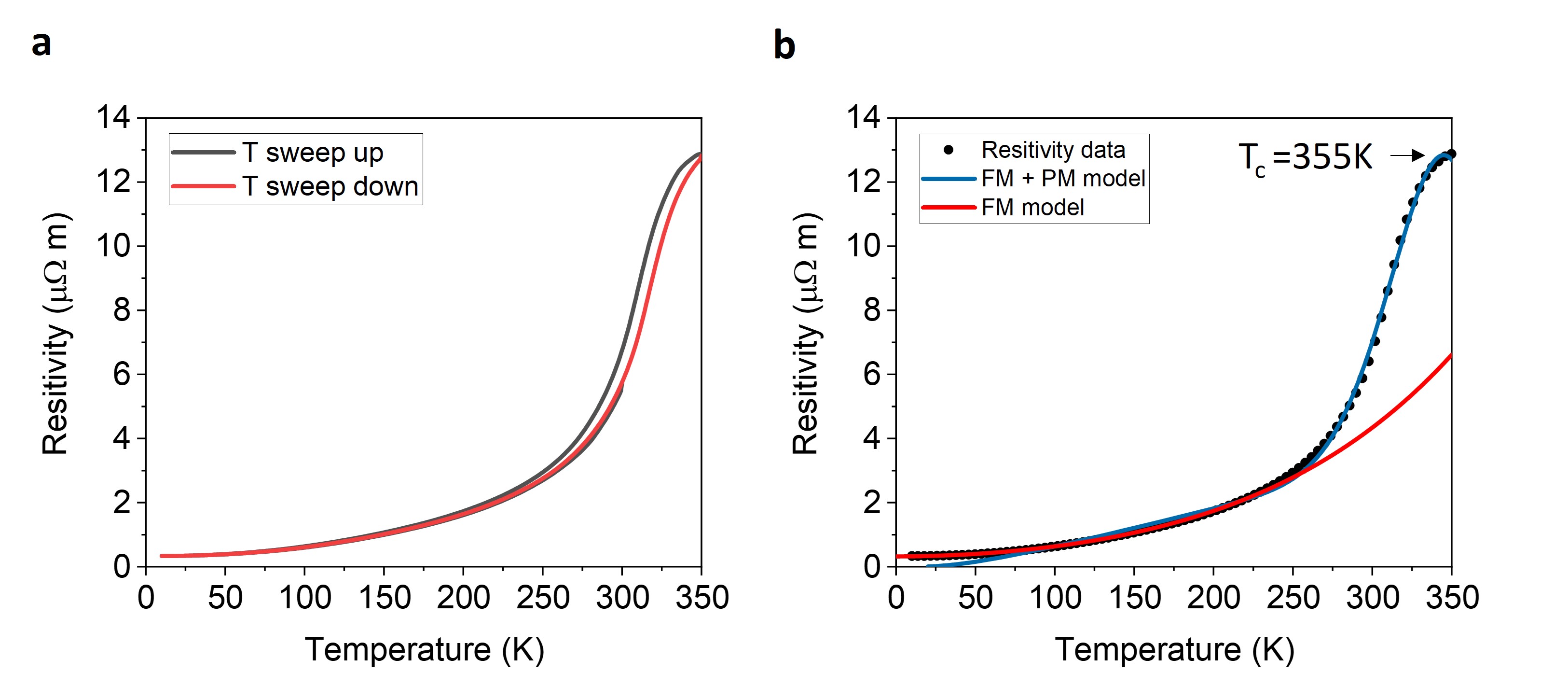}
            \caption{\textbf{Resistance vs temperature curves.} (\textbf{a}) Resistance vs temperature of \SI{13}{} u.c. LSMO thin film with resistance hysteresis observed between the upward (up) and downward (down) temperature sweeps. (\textbf{b}) Fitting a percolation model of resistivity of phase-separated ferromagnetic and paramagnetic areas (orange curve) \cite{Li2002CompetitionManganites} and a purely ferromagnetic model (red curve)  \cite{Li2002CompetitionManganites}.}
            \label{fig:resistance}
\end{figure*}

The distribution of the dI/dV map of \textbf{Fig. \ref{fig:STM_STS}d} indicates a bimodal distribution, shown in \textbf{Fig. \ref{fig:STM_STS}e}. A broader distribution of the blue areas is observed compared to the red ones. This distribution is an indication of chemical disorder caused by the local excess Sr and oxygen vacancies discussed previously, which thus still persist in the relatively flat surface regions. A possible effect of this disorder are
magnetic inhomogeneities due to the local variation in chemical composition. Such an magnetic inhomogeneity is important to consider because of the spin-ordering effect in the OER \cite{VanDerMinne2023DomainSplitting}.

To further verify this, we attempt to correlate a possible microscale
magnetic inhomogeneity with bulk transport measurements, which has been done before for manganites \cite{Miao2020DirectManganites, Baldini2015OriginManganite} and relates percolation theory \cite{NanPHYSICSMATERIALS} to the emergence of a nonlinear response such as magnetoresistance. In our work, we observed two nanoscale LDOS phases. In the percolation theory of phase separation, this electronic disorder introduces hysteresis or memory effects in R(T) and M(T) measurements. We performed the former, and the results in \textbf{Fig. \ref{fig:resistance}a} indicate such a hysteresis between the cooling-down and warming-up cycles. The temperature dependence of the resistance can be regarded as a system in competition between ferromagnetic (FM) metallic areas, with a certain spin correlation length, and paramagnetic (PM) areas with a semiconducting nature. Both regions can be identified by a varying spin-correlation length. For Ca-doped manganites, a phenomenological model was developed in Ref. \cite{Li2002CompetitionManganites} that describes the temperature dependence of the resistance as a percolation between a volume fraction of the PM and FM areas. Here, we fit the same model for LSMO, similar to the work in Ref. \cite{Yin2016StraininducedFilms}. In this model, the metallic conductivity is described with electron-electron interaction by a $T^2$ term and electron-spin fluctuation scattering by $T^{4.5}$ \cite{Li2002CompetitionManganites, Yin2016StraininducedFilms}. Therefore, the FM resistance can be given as $\rho_{FM}(T) = \rho_0 + AT^2 + BT^{4.5}$. For high temperatures, above $T_c$, the resistance can be described using a magnetic polaron picture \cite{Li2002CompetitionManganites} represented as $\rho_{PM}(T) = CT \text{exp}(\frac{E_g}{k_bT}$), with $E_g$ the activation energy and $k_b$ the Boltzmann constant. We fitted this phenomenological model to the R(T) data and a reasonable agreement is displayed in \textbf{Fig. \ref{fig:resistance}b}. For intermediate- and high-temperature regions, the fit agrees with the data, and for the low-temperature regime, a deviation is observed. We fitted the low-temperature regime with only the FM model, and a good agreement was reached, as indicated by the red line in \textbf{Fig. \ref{fig:resistance}b}. On the basis of these results, we conjecture that the observed structural and/or electronic phase separation of LSMO epitaxial thin films is likely correlated to the macroscopic film resistive behavior, and complex FM and PM phase separation occurs, especially at elevated temperatures.  

\section{Conclusion}

The combination of  XRD and RXR characterization and SPM imaging revealed a complex surface and electronic morphology of LSMO thin films grown on STO by PLD. Using element-specific RXR, we revealed a gradient in the chemical composition of the film that showed a stochiometric bulk but a non-stochiometric top layer of around \SI{1}{\nano\meter} with an excess of Sr, oxygen vacancies, formation of Mn$^{2+}$ valence states, and a La deficiency. This chemical disorder was further studied on the nanoscale with SPM where we revealed a staggered growth on the surface of the films consisting of nanoscale stacked islands and composed of spherical features. The formation of staggered growth was described by a two-step growth model driven by an Ehrlich-Schwoebel barrier on the surface and a reconstruction of the bulk yielding high crystallinity. We correlated the staggered growth with the formation of surface EPS driven by non-stochiometry and macroscale magnetic inhomogeneities. For applications such as electrochemistry correlated to the fundamental theoretical understanding of the complex oxides, careful consideration of the surface morphology and electronic structure is important. Therefore, the results presented in this work of the disentanglement between the bulk and surface properties of LSMO make it a tantalizing model system for evaluating the relationship between the surface and bulk spin ordering in electrocatalysis \cite{Xu2023MagneticCatalyst, Ren2021SpinpolarizedField, Li2019OptimizedReaction, VanDerMinne2023DomainSplitting}.

\section{Methods}

\subsection{PLD and RHEED}
Thin films of La$_{0.67}$Sr$_{0.33}$MnO$_3$ were fabricated using PLD and a stochiometric target from SurfaceNet. The vacuum system had a base pressure of \SI{5e-8}{\milli\bar} and was equipped with an in situ RHEED and a KrF excimer laser of \SI{248}{\nano\meter}. B-site terminated and step-terraced SrTiO$_3$ (001) substrates were purchased from CrysTec GmbH or Shinkosha Co. Ltd. The laser fluence was set to \SI{2.0}{\joule\per\square\centi\meter} and the frequency used was \SI{1}{\hertz}. The oxygen pressure was \SI{0.266}{\milli\bar} and the substrate temperature was \SI{750}{\celsius}. The distance between the sample and the target was \SI{5}{\centi\meter} and a rectangular mask was used to obtain a laser spot size of \SI{2.24}{\square\milli\meter}. The targets were pre-ablated at \SI{10}{\hertz} to remove any possible surface contamination. After deposition, the samples were slowly cooled with a rate of \SI{25}{\celsius}/min inside the PLD at \SI{100}{\milli\bar} oxygen pressure.

\subsection{XRD}
X-ray diffraction and reflectivity measurements were performed using a Bruker D8 Discover diffractometer with Cu-K$\alpha$ radiation and an Eiger 2 R 500K area detector. A Ge (022) monochromator was used for diffractograms and a collimator of \SI{1}{\milli\meter} diameter was used to obtain Cu-K$\alpha$1 radiation and shape the incident beam. The detector was operated in 0D mode with a small region of interest (\SI{975}{\micro\meter} x \SI{4575}{\micro\meter}) during the $2\Theta$-$\omega$ scans. To obtain the RSM, the detector was kept stationary while operating in 1D mode while an omega rocking curve was performed. Reflectivity measurements were made using a \SI{0.1}{\milli\meter} slit and a \SI{1}{\milli\meter} collimator to shape the incident beam. The detector was operated in 0D mode with a small region of interest (\SI{975}{\micro\meter} x \SI{4575}{\micro\meter}). Comparison of the measured and expected diffractograms of the LSMO film peaks around the STO 002 peak in a $2\Theta$-$\omega$ scan of the XRD spectra was performed using InteractiveXRDFit software \cite{Lichtensteiger2018InteractiveXRDFitHeterostructures}. More in-depth simulations of the XRD diffractogram in this study were carried out using the GID SL program at Sergei Stepanov's X-Ray Server \cite{Stepanov2004XrayScattering}. The parameters used to vary the crystallinity of the layer are the Debye-Waller like factors ($w_0$ and $w_h$), which form a correction to scattering and absorption factors based on the crystallinity of the material \cite{Stepanov2004XrayScattering}. Its value ranges from 0 to 1, where 0 is an amorphous layer and 1 corresponds to a perfect crystal without defects. For the more crystalline layer, the values of $w_0$ and $w_h$ were taken as \SI{0.85}{} to include a small number of defects and amount of disorder, typically present in PLD grown films. For the lower crystalline layer, a value of \SI{0.75}{} generated the best fit. A value of 1 was used for $w_0$ and $w_h$ for the crystalline substrate. The roughness values used were equal to \SI{2}{\angstrom} for the substrate and \SI{4.5}{\angstrom} for the film, in accordance with the roughness measured using AFM.

\subsection{RXR}
\subsubsection{Data Acquisition}
Data acquisition of the Resonant X-ray reflectometry (RXR) was done at the Resonant Elastic and Inelastic X-ray Scattering (REIXS) beamline of the Canadian Light Source (CLS) in Saskatoon, Canada \cite{Green2020ResonantHeterostructures}. A flux of \SI{5e12}{} photons per second with a photon energy resolution $\Delta$E⁄E of $\sim$10$^{-4}$ was used for the measurements. To give the desired linear and circular polarizations, an elliptical polarizing undulator was used. Measurements were done at a temperature of \SI{300}{\kelvin} under a base pressure of \SI{1e-9}{\milli\bar}. The reflection geometry scans were made possible by an in-vacuum 4-circle diffractometer after the samples were aligned with their surface normal in the scattering plane. Measurements were done at several resonant photon energies at different resonances: Ti L2,3 ($\sim$450-470 eV), Mn L2,3 ($\sim$635-660 eV), and La M4,5 ($\sim$830-860 eV), along with multiple nonresonant photon energies. The measurements were done in specular reflection geometry.  Circular-dichroic magnetic measurements were performed under a homogeneous \SI{0.6}{\tesla} field that was produced by inserting a permanent magnet array into the sample environment. This aligned the magnetization in both the $xy$-plane of the film and the scattering plane of the measurement. Dichroic measurements were done only at the resonant photon energies of the Mn L2,3 resonance. A photodiode was used to detect the reflected beam intensity, with the response function of the photodiode determined by directly measuring the synchrotron beam. All measured data were normalized by the incident beam flux and the response function to obtain quantitative reflectivity spectra. The full RXR dataset is presented in Ref. [\cite{VanDerMinne2023DomainSplitting}].

\subsubsection{Modelling}
Global Optimization of Resonant X-ray Reflectometry (GO-RXR), a software package recently developed by the QMaX group at the University of Saskatchewan was used for modelling of the RXR data. We used tabulated atomic form factors for nonresonant energies \cite{Henke1993XRay1-92} and measured x-ray absorption for the construction of resonant scattering tensors for elements Ti, Mn, and La. For Mn we differentiated the resonant scattering tensors for Mn$^{2+}$ and for Mn in stochiometric LSMO (implemented as a weighted linear combination of Mn$^{3+}$ and Mn$^{4+}$ scattering tensors corresponding to Mn$^{3.3+}$). A slab model was used to determine the optical and magneto-optical profiles. Such a model is made up of parametrized layers with defined elements, oxidation states, roughnesses, thicknesses, and densities. By modeling the film at the u.c. level using the model parameters we construct an element-specific discrete depth-dependent density profile. This so-called u.c. model fixed the thickness of each layer with a value that corresponds to the in-plane lattice constant of a theoretical LSMO/STO heterostructure and models the roughness as a step function. A density profile of the 13 u.c. LSMO samples is determined by optimizing the density of the elements and oxidation states present in each u.c. layer. First, an approximation of the density profile, along with the form factors, is used to determine the expected energy- and depth-dependent optical profile. The optical profile is then used to simulate the reflectivity for a given energy, reflection angle, and polarization. Finally, we optimized the density by fitting the simulations to the experimental data. The concentration of Sr and La is fixed to 3:7 throughout the bulk of the film (from the target stoichiometry) to reduce the parameter set; however it is allowed to vary at the interface and surface of the film. The nonmagnetic elemental density profile of Mn was first determined by optimizing the parameters against an extended sigma-polarized experimental dataset.

\subsection{Scanning probe microscopy}
\subsubsection{UHV atomic force microscopy}
Atomic force microscopy was performed with a Scienta Omicron GmBH UHV VT-SPM operating in ultra-high vacuum with a base pressure of \SI{e-10}{\milli\bar}. A self-sensing quartz tuning fork (AB38T) was customized as a two-prong force oscillating sensor \cite{Verhage2023SwitchablemagnetisationSensor} with a Q-factor of \SI{15000}{}.  Using a cleaved silicon wafer, a pyramidal-shaped tip is formed with a radius of about \SI{30}{\nano\meter}. Frequency-Modulation AFM using constant frequency shift feedback was used using a Phase-Lock Loop (PLL). The tip oscillated with an amplitude of \SI{1}{\nano\meter}. The ultrasharp-tip AFM sensor was fabricated by using a diamond tip (AdamaProbe). This tip was glued to one prong of the self-sensing tuning fork using EPOTEK E2101 UHV compatible two-component epoxy. AFM was operated in FM-AFM mode. The Q-factor of the sensor was around \SI{40000}{}. The tip oscillated with an amplitude of \SI{1.5}{\nano\meter}.
Post-processing of the AFM data was performed with Gwyddion software \cite{Necas2012GwyddionAnalysis}. Images were line aligned using median of differences and plane-leveled using mean subtraction. 

\subsubsection{Ambient atomic force microscopy}
Contact-AFM was performed using a Nanosensor NCH-PPP Si cantilever on a Veeco Dimension III microscope in ambient. Imaging was carried out with a normal force of \SI{5}{\nano\newton} with a scan speed of \SI{1} lines per second (512 lines/512 pixels). The friction image was obtained by registering the lateral movement of the cantilever detected by the 4-quadrant photodetector. 

\subsubsection{UHV scanning tunneling microscopy and scanning tunneling spectroscopy}
Scanning Tunneling Microscopy was performed with a Scienta Omicron GmbH UHV VT-SPM operating in ultra-high vacuum with a base pressure of \SI{e-10}{\milli\bar}. STM tips were mechanically cut from PtIr wire. The bias was applied to the tip and the sample was grounded. The imaging was performed in constant current mode. For dI/dV mapping and spectroscopy, an Oxford Instruments lock-in amplifier with an alternating voltage of \SI{100}{\milli\volt} applied at \SI{4.1}{\kilo\hertz} was used. Post-processing of the STM data was performed with Gwyddion software \cite{Necas2012GwyddionAnalysis}. The images were line-aligned using median of differences and plane-leveled using mean subtraction.

\begin{acknowledgement}
Financial support from the Eindhoven University of Technology is acknowledged. Support from the University of Twente in the framework of the tenure track start-up package, the Natural Sciences and Engineering Research Council of Canada (NSERC) Discovery Grant program, and the NSERC CREATE to INSPIRE program are gratefully acknowledged.
\end{acknowledgement}

\section*{Author contributions}
M.V., E.M, C.B., K.F. conceived and designed the experiments. E.M. synthesized the samples and performed RHEED during synthesis. E.M. and E.K. performed and analyzed x-ray diffraction measurements. L.K., R.S. and R.G. performed, modeled, and analyzed resonant X-ray reflectivity measurements. M.V. performed and analyzed scanning tunneling microscopy and spectroscopy mapping. M.V. and E.M. performed and analyzed physical property measurements. G.K. advised on the surface analysis and growth mode. C.B. and K.F. supervised the research. M.V. and E.M. wrote the manuscript with contributions from all authors. All authors have given their approval to the final version of the manuscript. 

\newpage
\section{Supplementary}

\subsection{S1}
\subsection{S2}

\newpage
\subsection{S3}

\subsection{S4}

\newpage
\subsection{S5}

\newpage

\bibliography{referencesLSMO.bib}

\end{document}